\documentstyle[graphics,psfig,12pt]{article}
\begin{document}

\noindent
\section*{Detection of magnetic dipole lines of \hbox{Fe\,{\sc xii}} in the
ultraviolet spectrum of the dwarf star $\epsilon$~Eri}\vskip0.10in
\noindent
{\bf C. Jordan$^*$, A. D. McMurry$^+$, S. Sim$^*$ \& M. Arulvel$^{\dag}$} 
\vskip0.10in
\noindent
* Department of Physics (Theoretical Physics), University of Oxford, 1
Keble Road, Oxford, OX1 3NP, UK \vskip0.10in
\noindent
+ Institute of Theoretical Astrophysics, University of Oslo,  P.O. Box
1029, Blindern, N-0315 Oslo, Norway \vskip0.10in
\noindent
\dag~St. Catherine's College, Oxford, OX1 3UJ, UK \vskip0.10in
\noindent
----------------------------------------------------------------------------
\vskip0.10in
\noindent

\section*{Abstract}
We report observations of the dwarf star $\epsilon$~Eri (K2~V) made with the 
Space Telescope Imaging Spectrograph (STIS) on the Hubble Space Telescope 
({\it HST}). The high sensitivity of the STIS instrument has allowed
us to detect the magnetic dipole transitions of \hbox{Fe\,{\sc xii}} at
1242.00~\AA~and 1349.38~\AA~for the
 first time in a star other than the Sun. The width of the stronger line at 
1242.00~\AA~has also been measured; such measurements are not possible
for the permitted lines of \hbox{Fe\,{\sc xii}} in the extreme
ultraviolet. To within the accuracy of the measurements, the \hbox{N\,{\sc v}}
 and the \hbox{Fe\,{\sc xii}} lines occur at their rest wavelengths. Electron
densities and line widths have been measured from other transition region
lines. Together, these can be used to investigate the non-thermal energy 
flux in the lower and upper transition region, which is useful in 
constraining possible heating processes. The
\hbox{Fe\,{\sc xii}} lines are also present in archival STIS spectra of 
other G/K-type dwarfs.   

\section*{Keywords}
stars: individual ($\epsilon$ Eridani)- 
stars: late-type - stars: coronae - 
line: identification - line: profiles 

\section{Introduction}

Ultraviolet spectra of cool dwarf stars observed with the
International Ultraviolet Explorer ({\it IUE}) and the Goddard High Resolution
 Spectrograph (GHRS) on the Hubble Space Telescope ({\it HST}) show emission
lines formed in stellar chromospheres and transition regions at 
temperatures of less than $\sim 2 \times 10^5$~K. In order to observe
emission lines formed at higher temperatures, observations in the extreme 
ultraviolet or X-ray regions are normally required. Simultaneous 
observations in these different wavelength regions are difficult to obtain. 
We have observed the dwarf star $\epsilon$~Eri (HD~22049, K2~V) with the 
Space Telescope Imaging Spectrograph (STIS) on the {\it HST}. This instrument 
has unprecedented sensitivity and spectral resolution 
at wavelengths down to $\sim 1150$~\AA~(Woodgate et al. 1998). The in-orbit 
performance of the STIS is
 described by Kimble et al. (1998). In addition to the usual chromospheric
 and low transition region lines observed in this star with IUE (Ayres et al. 
1983), we have detected two forbidden
(magnetic dipole) transitions in \hbox{Fe\,{\sc xii}}, which are
formed at an electron
temperature ($T_e$) of $\sim 1.4 \times 10^6$~K. These lines were first 
identified by Burton, Ridgeley \& Wilson (1967) in a solar spectrum
obtained during a Skylark rocket flight, but until now they have not been
observed 
in another star. Their detection allows the high temperature part of the 
stellar atmosphere ($T_{e} \ge 10^6$~K) to be studied simultaneously with the
 regions below $\sim 2 \times 10^5$~K. This is important since the X-ray 
emission from cool dwarfs may vary with time depending on the number of active
 regions present. The activity cycle period of $\epsilon$~Eri is
probably around 5 years (Gray \& Baliunas 1995). In addition to the line 
fluxes, we have measured the width of the stronger line. It is not possible to
measure the widths of permitted
\hbox{Fe\,{\sc xii}} lines with current extreme ultraviolet and soft
X-ray instruments. Together  
 with spectroscopic measurements of the electron density, line widths are 
 important in constraining possible coronal heating mechanisms (see e.g. 
Jordan 1991).  We also discuss the wavelengths measured for the 
\hbox{N\,{\sc v}} and \hbox{Fe\,{\sc xii}} lines. 

$\epsilon$~Eri is a well observed star at optical wavelengths (see
Drake \& Smith 1993) and has a measured spatially averaged magnetic
field of 165~Gauss (R\"uedi et al. 1997). It has also been
observed with the Extreme Ultraviolet Explorer ({\it EUVE}) (Schmitt
et al. 1996; Laming et al. 1996)
and in our {\it ROSAT} programme (Philippides 1996). It has a faster
rotation rate than the Sun (11.68 days) (Donahue, Saar \& Baliunas 1996)
and has higher surface fluxes and a higher mean coronal temperature of
$\ge 2.5 \times 10^6$~K (Laming, Drake \& Widing 1996; Philippides
1996).  $\epsilon$~Eri is therefore ideal for studying the structure
and heating of stars more active than the Sun. The stellar parameters used
below are summarized in Table 1. 

We have searched the STIS archive for spectra of the active G/K-type dwarfs studied by Montesinos \& Jordan~(1993). Both \hbox{Fe\,{\sc xii}} lines are present in STIS spectra of 70~Oph~A (HD 165341) and $\kappa$~Cet (HD 20630). The 1242.00-\AA~line is present in $\xi$~Boo~A (HD 131156A) and $\chi^{1}$~Ori (HD 39587) , but the 1349.38-\AA~ line is barely detectable. These stars have been modelled by Jordan~et~al.~(1987) and/or Philippides~(1996).

\section{Observations} 

The observations of $\epsilon$~Eri were made on 2000 March 9. Spectra 
covering  the wavelength range from 1150~\AA~to 1700~\AA~and from 1650~\AA~to
 3100~\AA~were obtained using the CsI MAMA (Multi-Anode Microchannel
 Array) detector and the CsTe MAMA detector, respectively. The most
 recent data reduction software has been used, which accounts for
 light scattered between different orders. Two sections of the spectrum are 
shown in Figs. 1a and 1b. These are the result of summing two separate 
exposures, which are individually very similar. Fig. 2 shows the same region 
as in Fig. 1a, in the STIS spectrum of $\xi$~Boo~A. 

\subsection{Line identifications and wavelengths}

Fig. 1a shows the resonance lines of
\hbox{N\,{\sc v}} ($2s~^{2}S_{1/2} - 2p~^{2}P_{3/2,1/2}$) and also a weaker, 
broader line. Fig. 1b shows the region around 1350~\AA, which contains 
another weak, broad line. These two broader lines are identified as the 
$3s^{2}3p^{3}~^{4}S_{3/2} - 3s^{2}3p^{3}~^{2}P_{3/2,1/2}$ magnetic dipole 
transitions in \hbox{Fe\,{\sc xii}}. The observed wavelengths (which include 
the correction for the stellar radial velocity of 15.5~km~s$^{-1}$) 
are given in Table 2. These were obtained by using a single Gaussian fit to 
 the line profiles.
  The absolute wavelength accuracy of the STIS instrument is determined by the
 detector pixel size which corresponds to 0.014~\AA~at 1238.82~\AA~and 
0.015~\AA~at 1349.40~\AA. Detector noise may in practice prevent this 
accuracy being achieved. Table 2 also gives the laboratory wavelengths of the 
 \hbox{N\,{\sc v}} lines taken from Kelly (1987). The wavelength
interval between these lines should be measurable to  
$0.007$~\AA. In both the individual spectra and the combined spectrum
our measured interval is slightly larger. The wavelength of the 
\hbox{Fe\,{\sc xii}}$^{4}S_{3/2} - ^{2}P_{1/2}$ transition has been 
determined from laboratory
observations of extreme ultraviolet lines (Jup\'en, Isler \& Tr\"abert 1993), 
but this has not proved possible for the $^{4}S_{3/2} - ^{2}P_{3/2}$ 
transition.
Instead, the solar wavelength of this transition is given (Sandlin et al. 
1986). Other solar measurements, often based on the assumption that lines of 
neutral atoms are unshifted, give the same value. However, the value adopted
differs from that given by Kelly (1987) by $- 0.03$~\AA. 

From Table 2 it can be seen that the 
\hbox{N\,{\sc v}} and \hbox{Fe\,{\sc xii}} lines are observed at their 
accepted rest wavelengths to within 0.018~\AA~(or 4.3~km~s$^{-1}$). Also, there
are no shifts in wavelength with respect to the optically thin lines of 
\hbox{O\,{\sc i}} at 1355.598~\AA~and 1358.512~\AA, to within the wavelength
accuracy possible. The \hbox{N\,{\sc v}} lines have some excess 
intensity in their wings above that derived from a single Gaussian fit, which
 will be discussed in future work. 

Many other forbidden lines of highly ionised species are observed in
 the solar uv spectrum (Jordan 1972; Sandlin, Brueckner \& Tousey 1977; 
Feldman \& Doschek 1977), but no others appear in the spectrum of 
$\epsilon$~Eri. The only other highly ionized forbidden line observed
in the ultraviolet spectra of stars other than the Sun is a transition of 
\hbox{Fe\,{\sc xxi}} at 1354.08~\AA, which was first observed in a spectrum of
 the dM0e star AU~Mic (Maran et al. 1994) obtained with the GHRS on the 
{\it HST}. 

\subsection{Line fluxes and widths}
The fluxes observed at the Earth in the \hbox{N\,{\sc v}} and 
\hbox{Fe\,{\sc xii}} lines and the line widths (full-width at
half-maximum) are also given in Table 2. These were also obtained using a 
single Gaussian fit to the line profiles, supplemented by measurements of the
total line flux above the local continuum.

To check whether or not $\epsilon$~Eri was unusually active at the time of 
observation, the fluxes in the \hbox{N\,{\sc v}} lines have been
compared with those observed with {\it IUE} (Jordan et al. 1987). The 
\hbox{N\,{\sc v}} lines were too weak to observe at high resolution with 
{\it IUE}. Their combined flux, plus the flux in the \hbox{Fe\,{\sc xii}} 
1242.00-\AA~line, was measured from a low resolution 
spectrum and was only 16 percent smaller than the present value. This
suggests that the level of activity at the time of the STIS and {\it IUE}
observations was not significantly different.

\section{Analysis of line fluxes}

\subsection{Electron density and pressure} 

Early calculations of the ratios of the fluxes in lines between levels of the
 \hbox{Fe\,{\sc xii}} ground configuration showed that these can be 
sensitive to the
 electron density ($N_e$) (Gabriel \& Jordan 1975; Flower 1977). The CHIANTI 
data base (Dere et al. 1997 v3.01) contains up-to-date
 atomic data and has been used to calculate the ratio of the fluxes in
 the observed lines of \hbox{Fe\,{\sc xii}} as a function of $N_e$ and
 $T_e$. At $T_e = 1.4
 \times 10^6$~K, the temperature at which the line emissivity has its
 maximum value, the flux ratio F(1242.00~\AA)/F(1349.38~\AA) is
 approximately constant between electron pressures of $P_{e}/k_{B} = 7
 \times 10^{14}$ to $6 \times 10^{15}$~cm$^{-3}$~K and 
then increases rapidly with $P_e$. The observed ratio gives an
 electron pressure of $7.0 \times 10^{15}$~cm$^{-3}$~K, but the
 substantial uncertainty in the ratio gives only an upper limit of $7.0
 \times 10^{16}$~cm$^{-3}$~K.  
Measurements of the flux ratios for several transition region line pairs which
 are sensitive to $N_e$ lead to a mean electron pressure of around 
$6 \times 10^{15}$~cm$^{-3}$~K. (This work will be reported in a later
 paper.) Only the \hbox{Si\,{\sc iii}} density sensitive line at
 1892.03~\AA~was observable in the
 {\it IUE} high resolution spectrum of $\epsilon$~Eri. When the density
 sensitive emission measure of the \hbox{Si\,{\sc iii}} line was compared to
 the mean emission measure over the temperature range where this line
 is formed, an electron pressure in the range $5 \times 10^{15} \le 
P_{e}/k_{B} \le 1.5 \times 10^{16}$~cm$^{-3}$~K was found (Jordan et
 al. 1987), consistent with the present improved mean value. Density 
sensitive lines are also observed in the {\it EUVE} spectrum, although line 
blending is significant and the spectra have a low
 signal-to-noise. Using line ratios in \hbox{Fe\,{\sc xiv}},  Laming et
 al. (1996) found densities corresponding to electron pressures between 
$10^{15}$ and  
$3 \times 10^{15}$~cm$^{-3}$~K, significantly lower than the values
 we derive. From the same spectrum, Schmitt et al. (1996) found a
 density corresponding to $P_{e}/k_{B} = 3.5 \times
 10^{15}$~cm$^{-3}$~K, but with $1 \sigma$ errors, values in the range 
$10^{15}$ to $10^{16}$~cm$^{-3}$~K were possible. 

\subsection{Emission measures}
The observed flux in the \hbox{Fe\,{\sc xii}} 1242.00-\AA~line has been 
converted to the stellar surface flux using the stellar distance and radius 
given in Table 1. This surface flux has then been used to calculate
the emission measure locus ($\int N_{e} N_{H} dh$ as f($T_e$)- see
Jordan et al. 1987 for the method).
 Using the atomic data from CHIANTI and the relative
ion number densities of Arnaud \& Rothenflug (1985), the peak emissivity
occurs at $1.4 \times 10^6$~K (for \hbox{Fe\,{\sc xii}}, other ion population
calculations give similar results). The stellar photospheric abundance of 
iron ($N_{Fe}/N_{H} = 2.57 \times 10^{-5}$) (Drake \& Smith 1993) was 
adopted. With a pressure of $6 \times 10^{15}$~cm$^{-3}$~K and
assuming a plane parallel layer, the emission measure derived at 
$1.4 \times 10^6$~K is $8.7 \times 10^{27}$~cm$^{-5}$. 

This emission measure can be compared with that derived from the flux 
observed in the extreme ultraviolet lines of \hbox{Fe\,{\sc xii}}
around 195~\AA~by
Laming et al. (1996) in spectra obtained with the {\it EUVE}. However,
unlike the ultraviolet region, the extreme ultraviolet region
is subject to absorption by the hydrogen and helium continuua
in the intervening interstellar medium. The interstellar 
absorption by hydrogen is usually found by comparing the observed H Lyman
$\alpha$ profile with an estimated intrinsic stellar profile. Dring et 
al. (1997) find a column density of $n_H = 7.6~(\pm 1.2) \times
10^{17}$~cm$^{-2}$. A new model of the
chromosphere and transition region of $\epsilon$~Eri will be made to
improve the intrinsic stellar profile. Other estimates rely on general 
relations derived from large surveys of interstellar absorption based on the 
spectra of hot stars.  For $\epsilon$~Eri, the latter method gives estimates 
of  $n_H$  between $1.5 \times
10^{18}$~cm$^{-2}$ (Laming et al. 1996) and $7.1 \times 10^{17}$~cm$^{-2}$
(Philippides 1996), leading to transmission factors at 195~\AA~between
0.74 and 0.88, respectively. The difference between the
above transmissions is not large at 195~\AA~and we adopt the value of 0.88. 

For consistency we calculated the emission measure locus from the photon 
fluxes given by Laming et al. (1996) using the same method as that
adopted for
the uv lines, rather than using the value derived by these authors or by
Schmitt et al. (1996).
The emission measure found from the \hbox{Fe\,{\sc xii}} line at 195.12~\AA~is
then $2.6 \times 10^{27}$~cm$^{-5}$, a factor of about 3 smaller than
that derived from the forbidden line. The value derived from the 
emissivities of Brickhouse, Raymond \& Smith (1995) is the same to within
$0.09$~dex. This difference of a factor of 3 is larger than the
uncertainty in the interstellar absorption. 
It is possible that, owing to the low signal-to-noise, the
\hbox{Fe\,{\sc xii}} {\it EUVE} line flux has been underestimated, but
an underestimate 
of a factor of 3 seems unlikely. Although the electron pressure derived has an
 estimated uncertainty of around $\pm 40$~percent, over the density range of 
interest, the emission measures have little sensitivity to $N_e$. A factor of 3
error in the relative excited level populations also seems unlikely. In their 
analysis of the \hbox{Fe\,{\sc xii}} lines observed in spectra
obtained during the 1970 solar eclipse, Gabriel \& Jordan 
(1975) derived $^{2}P$ populations by using the mean quiet Sun temperature
and density to normalize their calculations. The $^{2}P_{1/2}$ populations 
that they found agree with those from CHIANTI to within 10
percent. Variations in the level of activity between the times of the
STIS and {\it EUVE} observations cannot be excluded as the source of the 
discrepancy.

The emission measure derived from the \hbox{N\,{\sc v}} lines is $5.1 \times
10^{26}$~cm$^{-5}$ at $1.8 \times 10^5$~K. 

\section{Non-thermal energy fluxes}

To illustrate the potential use of both line intensities and widths we
consider the non-thermal energy flux that could be carried in a wave
mode, as given by
\begin{equation}
F_{nt} = \rho <v_{T}^{2}> v_{prop}.
\end{equation}
Here $\rho$ is the gas density, $v_T$ is the turbulent velocity and 
$v_{prop}$ is the wave propagation 
velocity. The observed line width yields the most probable non-thermal 
velocity $\xi$ (see e.g. Jordan et al. 1987) and then 
$<v_{T}^2> = 3/2~\xi^{2}$. Assuming a wave flux that propagates at the
Alfv\'en velocity ($B (4 \pi \rho)^{-1/2}$), the non-thermal energy
flux can be found from the \hbox{N\,{\sc v}} and \hbox{Fe\,{\sc xii}}
lines. The values derived
are $4.2 \times 10^5 B$~erg~cm$^{-2}$~s$^{-1}$ from the \hbox{N\,{\sc v}} 
lines and 
$1.8 \times 10^5 B$~erg~cm$^{-2}$~s$^{-1}$~
from the \hbox{Fe\,{\sc xii}} 1242.00-\AA~line. If both sides of equation (1)
are multiplied by $A$, the area occupied by the magnetic field, then if there
is no energy dissipation until the corona is reached, 
one would expect the
same values of $F_{nt} A$ from the different lines, which is not the case.
 However, a varying
inclination of the mean magnetic field will affect the values of $\xi$ derived
 and not all the magnetic flux present in the transition region may
continue to the corona.

The energy flux upwards at $1.4 \times 10^6$~K, where the
\hbox{Fe\,{\sc xii}} lines are formed, can be compared
with the energy conducted down from the corona at this temperature.
The conductive flux is given by
\begin{equation}
  F_{c} = - \kappa T_{e}^{5/2} \frac{\mbox{d}T_{e}}{\mbox{d}h}
\end{equation}
where $\kappa = 1.1 \times 10^{-6}$~erg~cm$^{-1}$~s$^{-1}$~K$^{-7/2}$.
The temperature gradient can be written in terms of the local emission measure
 averaged over the typical temperature range of line formation
 ($\Delta \log T_{e} = 0.30$~dex) using $Em(0.3) \simeq Em(T_{e})/0.7$ and 
\begin{equation}
 Em(0.3) \simeq \frac{0.83 P_{e}^{2}}{\sqrt 2 \mbox{k}_{\mbox{\scriptsize B}}^{2} T_{e}} 
         \frac{\mbox{d}h}{\mbox{d}T_{e}}
\end{equation}
(see Jordan et al. 1997 for details). This gives 
$F_{c} = 3.5 \times 10^6$~erg~cm$^{-2}$~s$^{-1}$ at $1.4
\times 10^6$~K. The coronal radiation loss should also be included, but for
reasons of thermal stability should not exceed $F_c$.
Thus equating $F_{c}$ to the upwards energy flux
results in a magnetic flux density of 19~Gauss, to within a factor of two.
 This is an upper limit since the  \hbox{Fe\,{\sc xii}} 
emission measure is a spatially averaged value. Since the observed
spatially averaged surface magnetic field is 165~Gauss, not all 
the surface flux is required to extend to $1.4 \times 10^6$~K.

In future work the full range of line fluxes and widths observed will be used
 to make
a new model of the outer atmosphere of $\epsilon$~Eri and to investigate the 
energy balance and heating requirements in more
detail. The above numerical estimates are meanwhile illustrative. 

\section{Conclusions} 

The newly identified lines of \hbox{Fe\,{\sc xii}} offer the first
opportunity for simultaneous studies of the upper transition region/corona and the lower transition
region of moderately active dwarfs. Line fluxes, widths and shifts 
are useful in investigating processes related to the coronal heating.
  Although the spectra of $\epsilon$~Eri have the best signal-to-noise ratio, 
useful data can be extracted from the spectra of several other dwarfs.

\section*{Acknowledgements}

S. A. Sim acknowledges support as a PPARC funded D.Phil student. A. D. McMurry was supported by PPARC grant GR/K~98469 at the time when the observations were planned.

\section*{References} 

Arnaud M., Rothenflug R., 1985, A\&AS, 60, 425     \\
\noindent
Ayres T. R., Linsky J. L., Simon T., Jordan C., Brown A., 1983, ApJ, 274, 784
\\
\noindent
Brickhouse N. S., Raymond J. C., Smith B. W., 1995, ApJS, 97, 551\\
\noindent
Burton W. M., Ridgeley A., Wilson R., 1967, MNRAS, 135, 207 \\
\noindent
Dere K. P., Landi E., Mason H. E., Monsignori-Fossi B. C., Young
P. R., 1997, \\
\indent A\&AS, 125, 149 \\
\noindent
Donahue R. A., Saar S. H., Baliunas S. L., 1996, ApJ, 466, 384 \\
\noindent
Drake J. J., Smith G., 1993, ApJ, 412, 797 \\
Dring A. R., Linsky J. L., Murthy J., Henry R. C., Moos W.,
Vidal-Madjar A., \\
\indent Audouze J., Landsman W., 1997, 488, 760 \\  
\noindent
ESA, 1997, The Hipparcos and Tycho Catalogues, ESA SP-1200, Nordwijk \\
\noindent
Feldman U., Doschek G. A., 1977, JOSA, 67, 726  \\
\noindent
Flower D., 1977, A\&A, 54, 163   \\
\noindent
Gabriel A. H., Jordan C., 1975, MNRAS, 173, 397 \\
\noindent
Gray D.F., Baliunas S. L., 1995, ApJ, 441, 436 \\
\noindent
Jordan C., 1972, Sol. Phys. 21, 381 \\
\noindent
Jordan C., 1991, in {\it Mechanisms of Chromospheric and Coronal Heating},
eds, P. \\ 
\indent Ulmschneider, E. R. Priest, R. Rosner, Springer-Verlag, Berlin, p. 300
  \\
\noindent
Jordan C., Ayres T. R., Brown A., Linsky J. L., Simon T., 1987, MNRAS,
225, 903 \\
Jup\'en C., Isler R. C., Tr\"abert E., 1993, MNRAS, 264, 627 \\
\noindent
Kelly R. L., 1987, J. Phys. Chem. Ref. Data, 16 (Suppl. 1), 1 \\
\noindent
Kimble R. A. et al., 1998, ApJ, 492, L83 \\
\noindent
Laming J. M., Drake J. J., Widing K. G., 1996, ApJ, 462, 948 \\
\noindent
Maran S. P., et al., 1994, ApJ, 421, 800\\
\noindent 
Montesinos B., Jordan C., 1983, MNRAS, 264, 900\\
\noindent 
Philippides D.A., 1996, D.Phil. Thesis, University of Oxford \\
\noindent
R\"uedi L., Solanki S. K., Mathys G., Saar S. H., 1997, A\&A, 318,
429 \\
\noindent
Sandlin G. D., Brueckner G. E., Tousey R., 1977, ApJ, 214, 898 \\
\noindent
Sandlin G. D., Bartoe J.-D. F., Brueckner G. E., Tousey R., VanHoosier
M. E.,\\ 
\indent 1986, ApJS, 61, 801 \\
\noindent
Schmitt J. H. M. M., Drake J. J., Stern R. A., Haisch B. M., 1996, ApJ,
457, 882  \\
\noindent
Woodgate B. et al., 1998, PASP, 110, 1183 \\

\newpage

\section*{Tables}

\begin{table}[h]
\caption{Parameters for $\epsilon$~Eri} 
\begin{center}
\begin{tabular}{lccc}
 \hline
Distance$^a$  & Radius$^b$ & Radial Velocity$^c$ & Magnetic field$^d$  \\
 (pc)   &    ($R_{\odot}$)  & (km~s$^{-1}$)  & (Gauss) \\
\hline
3.218   &      0.792        &  15.5 &   165($\pm 30$)  \\
\hline
\end{tabular}
\end{center}

\noindent $^a$The Hipparcos catalogue (ESA 1997); $^b$Using angular
diameter 2.29 $\times 10^{-3}$~arcsec from Ayres et al. (1983);
$^c$SIMBAD data base; 
$^d$Spatially averaged value from R\"uedi et al. (1997). \\
\vspace{1cm}
\end{table}

\newpage 

\begin{table}[t]
\caption{The wavelengths, fluxes and widths of the \hbox{N\,{\sc v}} and 
\hbox{Fe\,{\sc xii}} lines}
\begin{center}
\begin{tabular}{lcccc}
\hline
Ion  &  Wavelength$^{a}$ & Wavelength$^{b}$   &   Flux at Earth & Line width
  (FWHM) \\
     &    \AA  & \AA    &   $10^{-14}$ erg~cm$^{-2}$~s$^{-1}$ &     \AA     \\
\hline
\hbox{N\,{\sc v}} & 1238.821 & 1238.828 & 9.4($\pm 0.3$) & 0.178($\pm 0.004$)
 \\
\hbox{N\,{\sc v}} & 1242.804 & 1242.822 & 4.6($\pm 0.2$) & 0.172($\pm 0.004$)
 \\
\hbox{Fe\,{\sc xii}} & 1242.00($\pm 0.01$) & 1241.992 & 0.98($\pm 0.07$) & 
0.21($\pm 0.04$) \\
\hbox{Fe\,{\sc xii}} & 1349.38($\pm 0.02$) & 1349.403($\pm 0.02$) & 
0.52($\pm 0.15$) & 
    -     \\
\hline
\end{tabular}
\end{center}

\noindent $^a$Wavelength derived from laboratory or solar observations - 
see text; $^b$Observed wavelength, corrected for the stellar radial velocity.
The wavelength error is $\ge \pm 0.007$~\AA, except for the 1349.38-\AA~line.
\end{table}

\newpage

\section*{}
\newpage

\section*{Figure captions}

\noindent
{{\bf Figure~1a}. The observed spectrum of $\epsilon$~Eri showing the 1238.82- and 1242.80-\AA~lines of \hbox{N\,{\sc v}} and the 1242.00-\AA~line of \hbox{Fe\,{\sc xii}}.}

\vspace{2cm}

\noindent
{{\bf Figure~1b}. The observed spectrum of $\epsilon$~Eri showing the 1349.38-\AA~line of \hbox{Fe\,{\sc xii}} and a line of \hbox{Cl\,{\sc i}} at 1357.55~\AA. Note that the flux scale is expanded by a factor of 2 compared to Figure~1a.}

\vspace{2cm}

\noindent
{{\bf Figure~2a}. The observed spectrum of 70~Oph~A showing the 1238.82- and 1242.80-\AA~lines of \hbox{N\,{\sc v}} and the 1242.00-\AA~line of \hbox{Fe\,{\sc xii}}.}

\vspace{2cm}

\noindent
{{\bf Figure~2b}. The observed spectrum of $\xi$~Boo~A showing the 1238.82- and 1242.80-\AA~lines of \hbox{N\,{\sc v}} and the 1242.00-\AA~line of \hbox{Fe\,{\sc xii}}.}

\newpage
\twocolumn

\section*{}

\begin{figure}
\centerline{\hbox{
\psfig{file=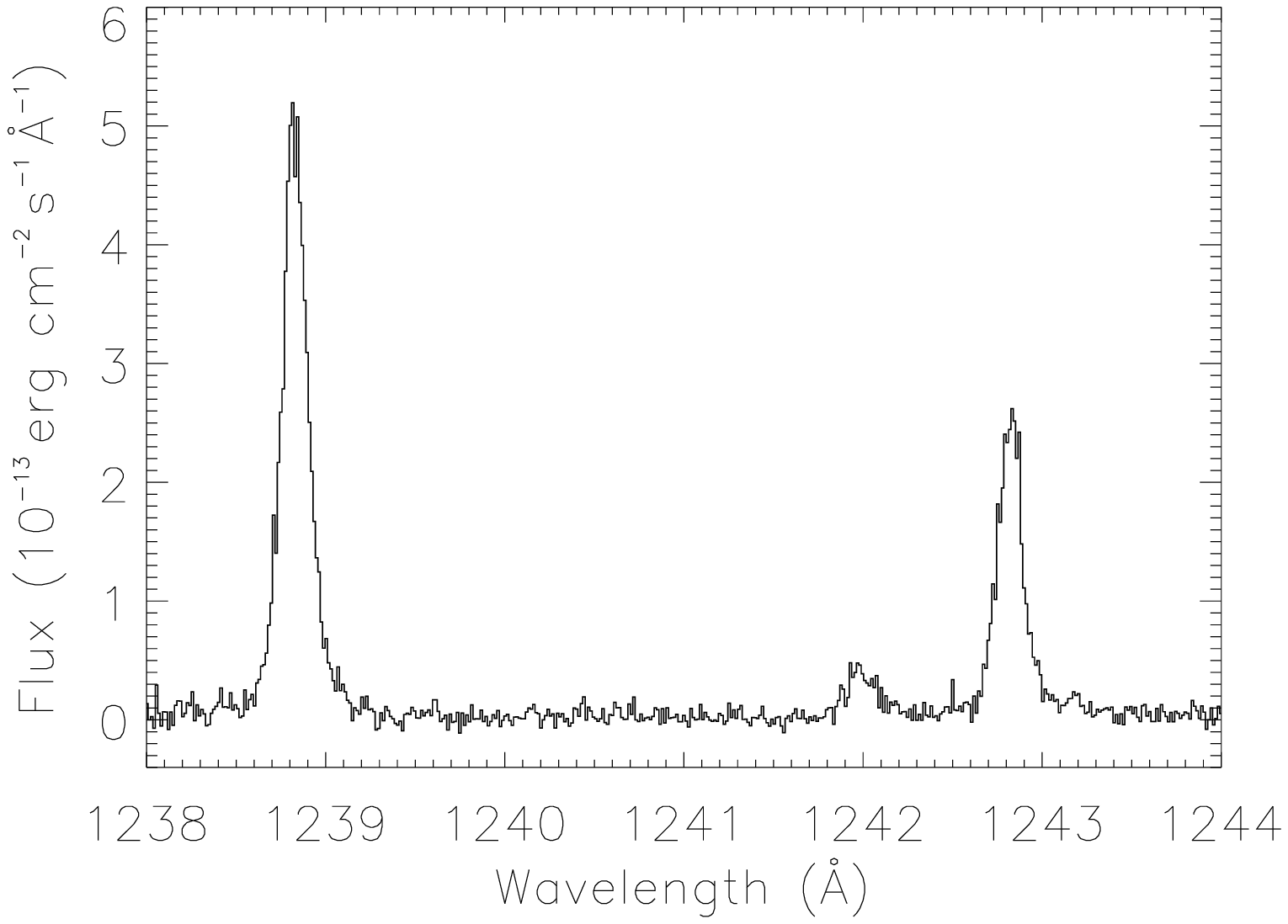,width=9cm,height=8cm}
}}
\begin{center}
{{\bf Figure~1a}.}
\vspace{2cm}
\end{center}
\end{figure}

\newpage
\section*{}

\begin{figure}
\centerline{\hbox{
\psfig{file=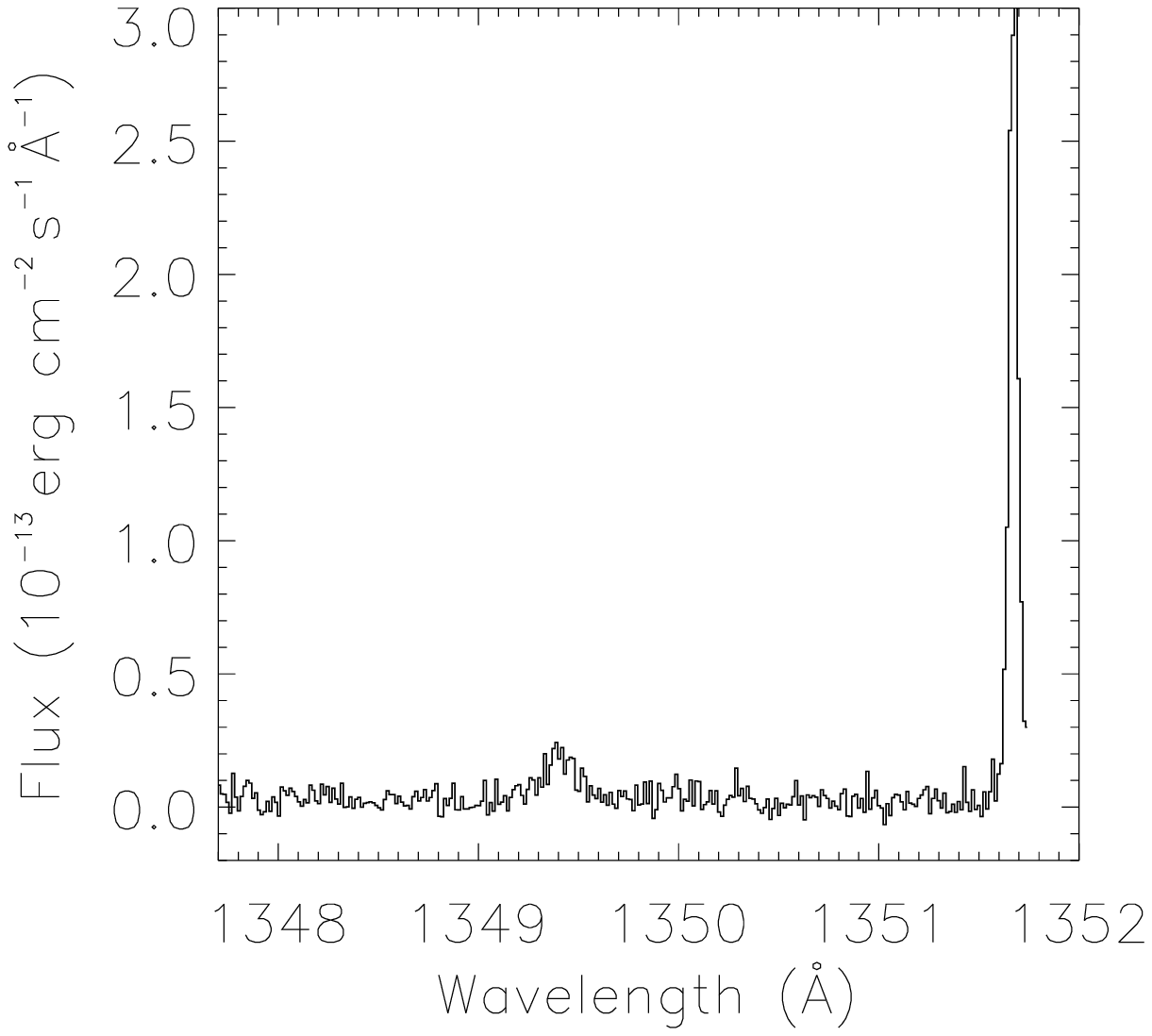,width=9cm,height=8cm}
}}
\begin{center}
{{\bf Figure~1b}.}
\vspace{2cm}
\end{center}
\end{figure}

\newpage

\section*{}

\begin{figure}
\centerline{\hbox{
\psfig{file=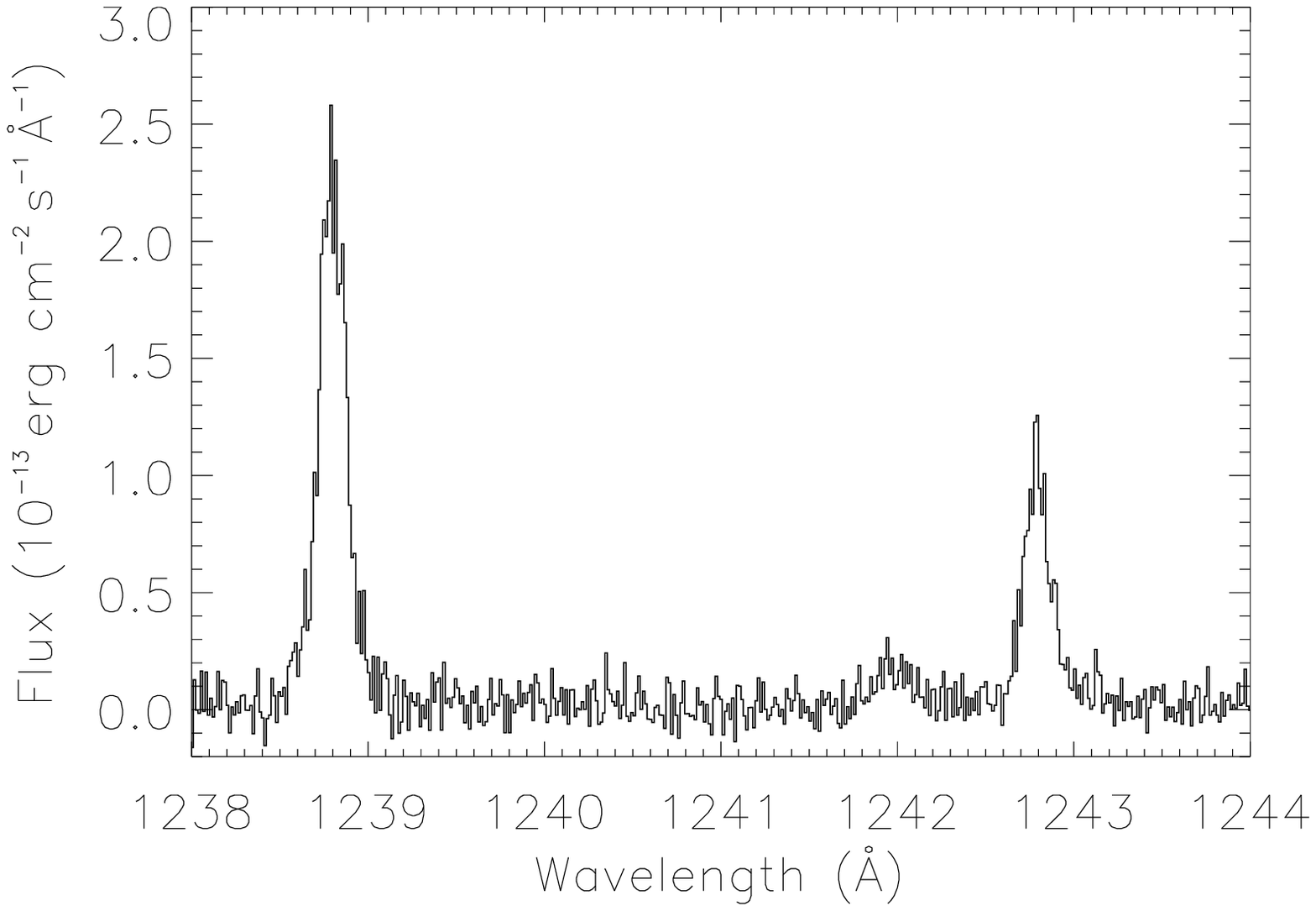,width=8.5cm,height=8cm}
}}
\begin{center}
{{\bf Figure~2a}.}
\end{center}
\end{figure}

\newpage

\section*{}

\begin{figure}
\centerline{\hbox{
\psfig{file=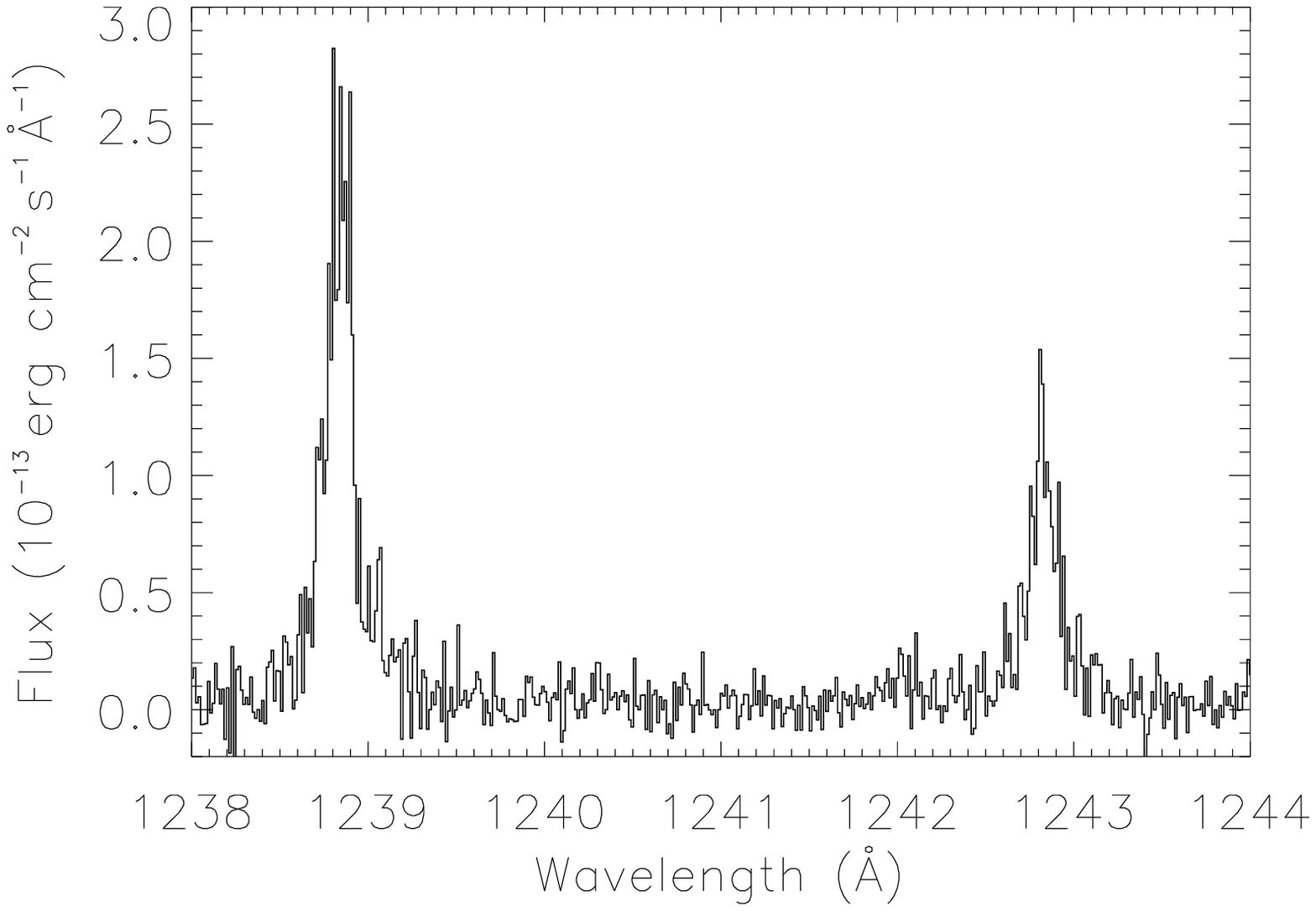,width=8.5cm,height=8cm}
}}
\begin{center}
{{\bf Figure~2b}.}
\end{center}
\end{figure}

\end{document}